\documentclass[journal=jacsat,manuscript=article,layout=twocolumn]{achemso}
\usepackage[T1]{fontenc}
\usepackage{xurl}
\usepackage{times}
\usepackage[english]{babel}
\usepackage[utf8]{inputenc}

\usepackage{fancyhdr}
\newcommand{\angstrom}{\mbox{\normalfont\AA}}

\usepackage[margin=2.54cm]{geometry}

\usepackage{lipsum}
\usepackage{mhchem}
\usepackage{amsmath}
\usepackage{graphicx}
\usepackage[colorlinks=true, allcolors=blue]{hyperref}

\title{Laser-Induced Topological Toggle Switching at Room Temperature in the van der Waals Ferromagnet \ce{Fe3GaTe2}}

\author{Charlie W. F. Freeman}
\affiliation{London Centre for Nanotechnology, University College London, London, WC1H 0AH, UK}
\alsoaffiliation{National Physical Laboratory, Teddington, TW11 0LW, UK}
\alsoaffiliation{Department of Electronic and Electrical Engineering, University College London, London, WC1E 7JE, UK}
\email{uceecwf@ucl.ac.uk}

\author{Woohyun Cho}
\affiliation{Department of Physics, Korea Advanced Institute of Science and Technology (KAIST), Daejeon 34141, Republic of Korea}

\author{Paul S. Keatley}
\affiliation{Department of Physics and Astronomy, University of Exeter, EX4 4QL, Exeter, UK}

\author{PeiYu Cai}
\affiliation{School of Physics and Astronomy, The University of Edinburgh, Edinburgh EH9 3FD, U.K.}

\author{Elton J. G. Santos}
\affiliation{Institute for Condensed Matter Physics and Complex Systems,
School of Physics and Astronomy, The University of Edinburgh, Edinburgh EH9 3FD, United Kingdom}
\alsoaffiliation{Donostia International Physics Center (DIPC), 20018 Donostia-San Sebastián, Spain}
\author{Robert J. Hicken}
\affiliation{Department of Physics and Astronomy, University of Exeter, EX4 4QL, Exeter, UK}
\author{H. Yang}
\affiliation{Department of Physics, Korea Advanced Institute of Science and Technology (KAIST), Daejeon 34141, Republic of Korea}

\author{Hidekazu Kurebayashi}
\affiliation{London Centre for Nanotechnology, University College London, London, WC1H 0AH, UK}
\alsoaffiliation{Department of Electronic and Electrical Engineering, University College London, London, WC1E 7JE, UK}
\alsoaffiliation{WPI-AIMR, Tohoku University, 2-1-1, Katahira, Sendai 980-8577, Japan}

\author{Murat Cubukcu}
\affiliation{London Centre for Nanotechnology, University College London, London, WC1H 0AH, UK}
\alsoaffiliation{National Physical Laboratory, Teddington, TW11 0LW, UK}
\email{m.cubukcu@ucl.ac.uk}

\author{Maciej Dabrowski}
\affiliation{Department of Physics and Astronomy, University of Exeter, EX4 4QL, Exeter, UK}
\email{m.k.dabrowski@exeter.ac.uk}

\date{\today}

\begin{document}

\maketitle   
\begin{abstract}

We demonstrate room-temperature nucleation and manipulation of topological spin textures in the van der Waals (vdW) ferromagnet \ce{Fe3GaTe2} through laser pulse excitation. By leveraging laser-induced heating and subsequent cooling, we access the skyrmion/bubble state at low fields and achieve toggle switching between two topological spin textures - skyrmion/bubble and labyrinth. Micromagnetic simulations reveal that this switching behaviour arises from laser-induced heating and cooling. Our findings highlight the potential of vdW ferromagnets for room temperature laser-controlled non-volatile memory storage applications. 
\begin{figure}
        \centering
        \includegraphics[width=0.8\linewidth]{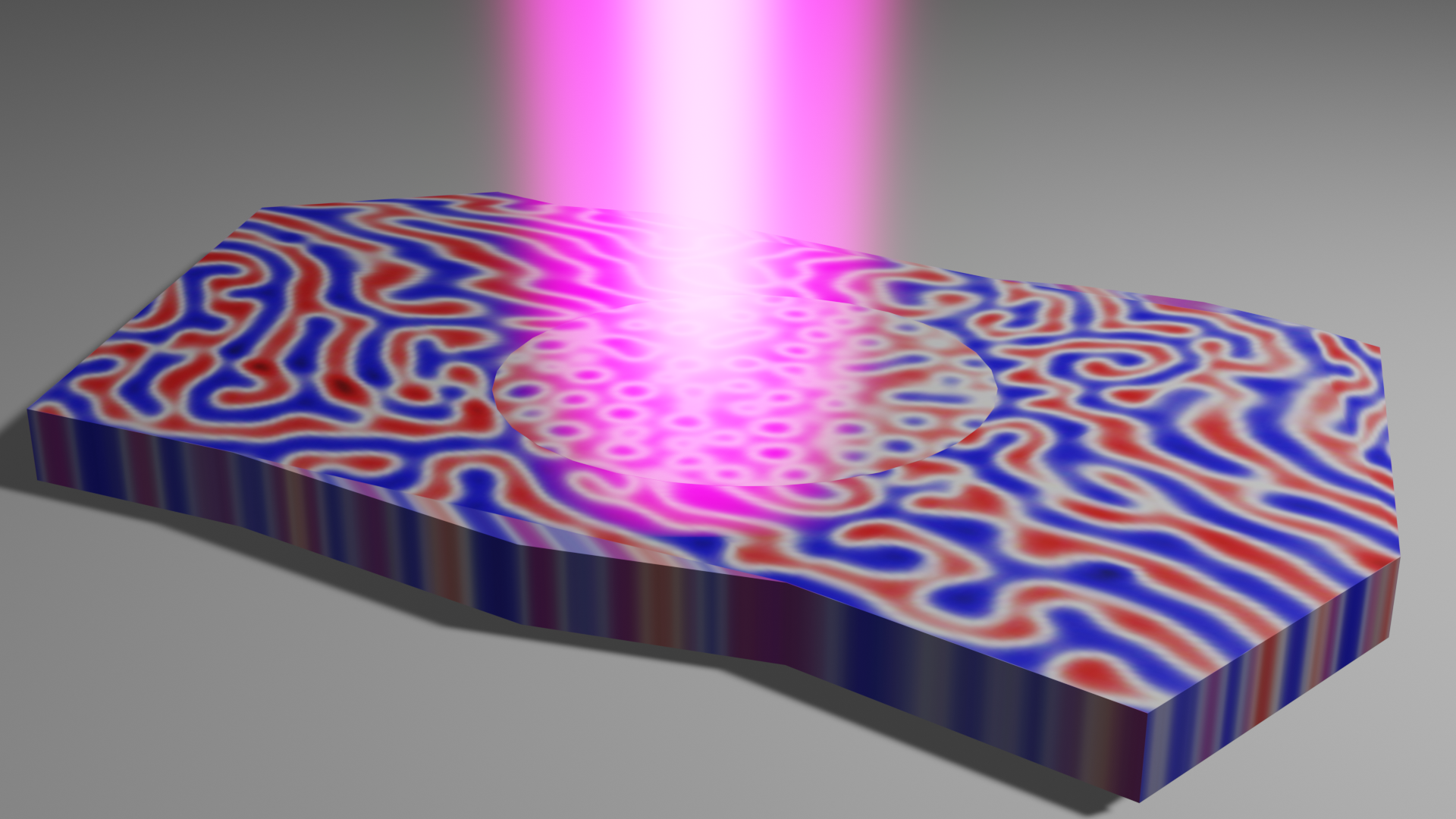}
    
    \end{figure}
\end{abstract}
\thispagestyle{fancy}
\setlength{\footskip}{15pt}

\section*{Introduction}
The discovery of stable, long-range magnetic order in single-layer vdW materials has stimulated a new interest in the study of layered magnetic materials, particularly for future spintronic applications \cite{gong2017discovery,huang2017layer,kurebayashi2022magnetism,jiang2021recent,gong2019two, wang2025configurable,burch2018magnetism,wang2022magnetic}. Most known vdW ferromagnets exhibit topological spin textures in their pristine state, due to the presence of an intrinsic Dzyaloshinskii–Moriya interaction (DMI) from broken inversion symmetry \cite{Zhou2025}. Recently, laser pulses have emerged as an effective tool for creating and manipulating these textures \cite{Berruto2018,Gerlinger2021, haldar2025all,Kern2025}. Through ultrafast demagnetisation \cite{Beaurepaire1996}, laser pulses can realise rapid heating and induce non-trivial topological states not accessible via standard field-cooling \cite{Kern2025,Khela2023}. For potential applications, deterministic toggle switching between states is essential, similar to all-optical switching (AOS) of uniform/monodomain magnetisation \cite{Lambert2014,Stanciu2007,Dabrowski2021}. While vdW magnets have demonstrated AOS \cite{Zhang2022a,Dabrowski2022} and optical control of topological textures \cite{Khela2023}, such studies have so far been restricted to cryogenic temperatures.

The synthesis of \ce{Fe3GaTe2} represents an important advancement toward above-room-temperature vdW systems, with a Curie temperature ($T_\mathrm{C}$) of 350~K - 370~K, thereby enabling the exploration of spintronic applications for vdW magnets under ambient conditions \cite{zhang2022above,lv2024distinct,hu2024room,ruiz2024origin,liu2024magnetic}. Spin structures such as skyrmions, antiskyrmions, and skyrmion bags have been already reported for \ce{Fe3GaTe2}, with demonstrations of zero-field skyrmion stability from above room temperature down to 100 K, when formed via field cooling procedures \cite{li2024room,zhang2024above,lv2024distinct,luo2025manipulation}. Further studies have explored the manipulation and formation of skyrmions through field cooling processes combined with stray fields from magnetic force microscopy tips \cite{mi2024real}, and the nucleation of skyrmions through ultra-fast laser writing \cite{li2024room} in non-stochiometric Fe$_{2.84\pm0.05}$GaTe$_2$, where ultrafast heating and subsequent quenching in an external field forms the skyrmion spin structure. However, laser-driven toggle switching between skyrmion/bubble and labyrinth domain states remains unexplored, particularly at room temperature.


In this study, we demonstrate room-temperature, low-field nucleation of topological spin textures through the application of laser pulses. The skyrmion/bubble phase is accessed via laser-induced heating followed by rapid quenching under external fields. Furthermore, we demonstrate toggle switching between two distinct topological spin textures, the skyrmion/bubble state and the labyrinth domain state through a combination of laser and magnetic field protocols, expanding the possibilities for topological spin texture control through laser manipulation. Micromagnetic simulations show that this behaviour can be explained by laser-driven heating and cooling, stabilising different phases depending on the applied field.

\section*{Results and Discussion}

\ce{Fe3GaTe2} has hexagonal structure with space group P6$_3$/mmc with the crystalline axis defined by $a = b = 3.9860 \;{\angstrom}, c = 16.2290 \;\angstrom, \alpha = \beta = 90°, \gamma = 120°$ \cite{zhang2022above}. In its bulk single crystal, \ce{Fe3GaTe2} has the saturation magnetisation (\( M_s) \) of 40.11 emu/g and large perpendicular magnetic anisotropy (PMA), $K_\mathrm{U}$, of -4.79x10$^5$ $J/m^3$ \cite{zhang2022above}. We employ wide-field Kerr microscopy (WFKM) in polar geometry to probe the domain structure, the magnetisation reversal and optically-induced topological spin textures.  

In Fig. \ref{fig1}a, we present representative WFKM images at selected magnetic field values. The saturation field at room temperature is found to be approximately 280 mT. The domain morphology vs field is typical of materials with PMA that undergo nucleation of labyrinth (or alternatively called stripe) domains as the field decreases (. The average width of the stripe domain is found to be 370 $\pm 90$ nm through ridge detection analysis at zero field and room temperature \cite{steger1998unbiased}.

\begin{figure*}[!ht]
    \includegraphics[width = 1\linewidth]{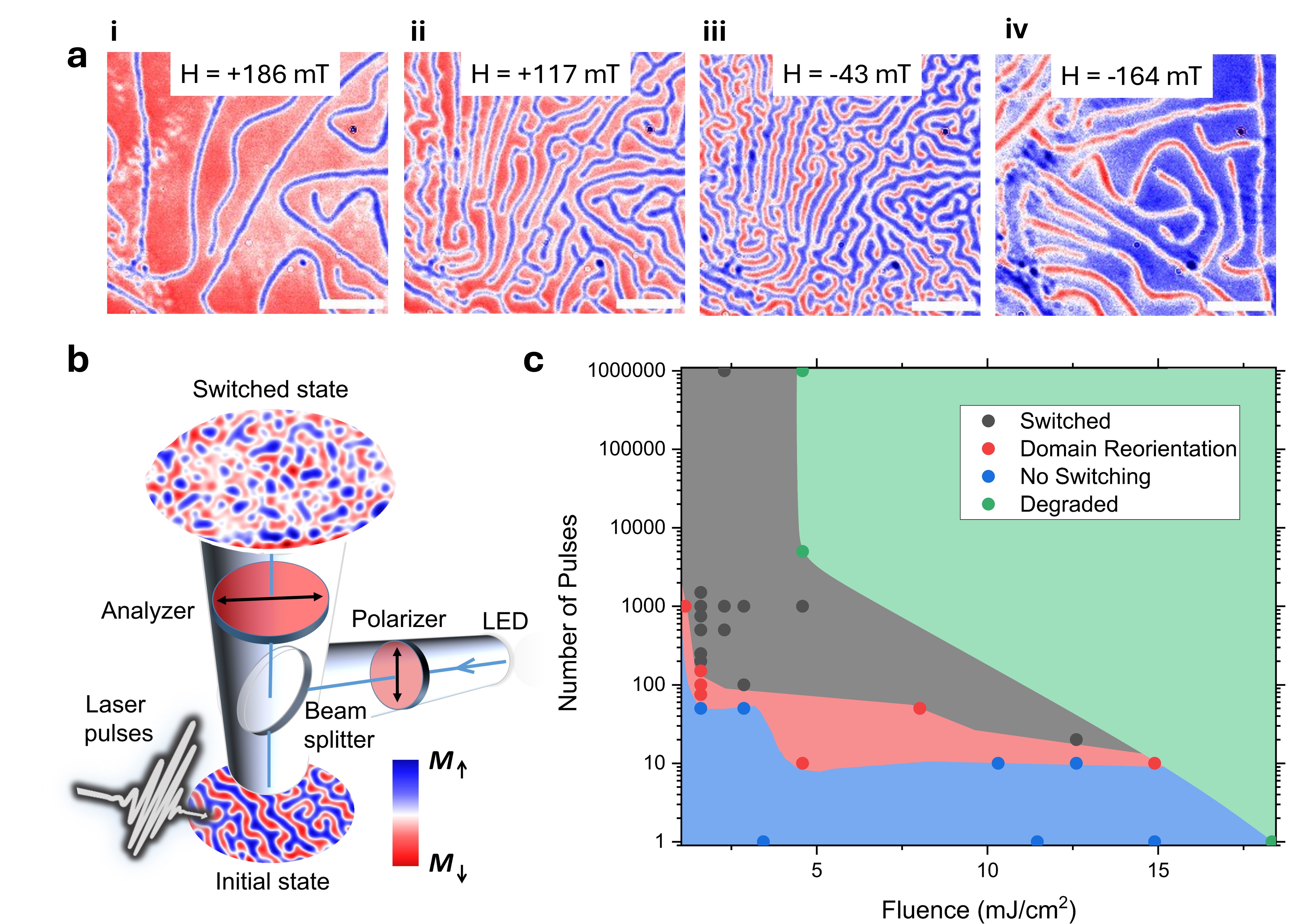}
    \caption{Magnetisation reversal and  laser-induced topological spin switching. (a) Wide-field Kerr microscopy (WFKM) images acquired at applied magnetic fields of (i) 186 mT, (ii) 117 mT, (iii) –43 mT, and (iv) –164 mT, showing magnetisation reversal. The scale bar is 5 $\mu$m.  (b) Schematic of the experimental setup for WFKM combined with external laser excitation. WFKM images show laser-induced switching at remanence from the initial labyrinth state to a skyrmion/bubble structure with optical pumping for 1s at 100 mW with a small +6 mT field applied. (d) Diagram showing the switching behavior as a function of fluence and the number of pulses, the regions are guides for the eye. The measurements were performed at room temperature.}
\end{figure*}
    \label{fig1}

\begin{figure*}
    \centering
    \includegraphics[width = 1\linewidth]{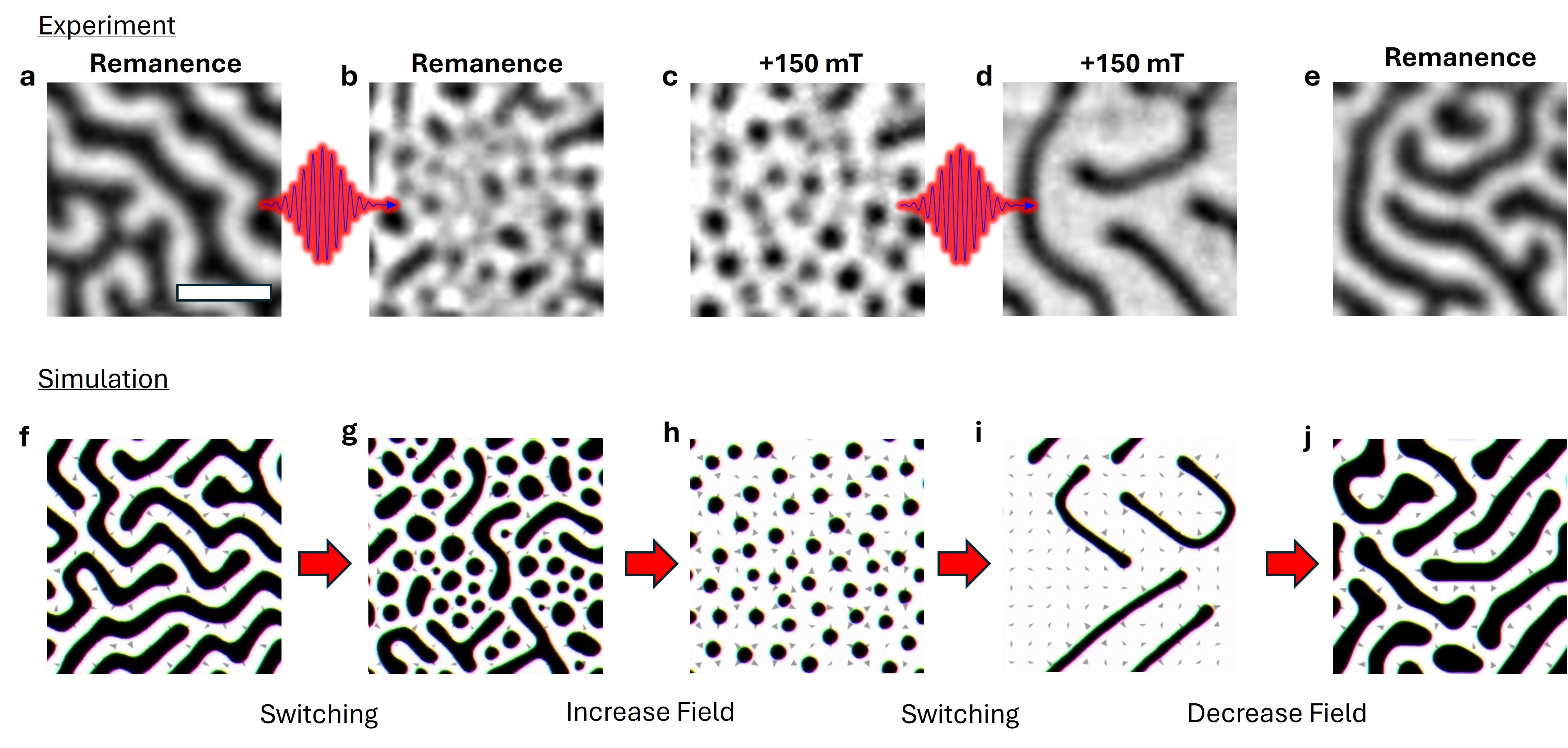}
    \caption{(a-e) Light-induced toggle switching between labyrinth and skyrmion/bubbles states via WFKM. Starting from an initial remanent labyrinth domain structure (a), laser pulses  for 1s at 100 mW induce a transition to a magnetic configuration dominated by the skyrmion/bubble state (b), which becomes even more clearly identifiable after applying a magnetic field of +\,150 mT (c). Subsequent exposure of the skyrmion/bubble state to laser pulses triggers another topological spin switching, leading to a transition back to the labyrinth state (d). Removing the magnetic field (e) restores the system to the same state as in (a). The scale bar has length 2 $\mu$m. (f-j) Micromagnetic simulations replicate the sequence shown (a-e). The simulation size is 1 $\mu$m $\times$ 1 $\mu$m $\times$ 50 nm. The field used for switching in (g) is +11 mT and 140 mT (i). }
    \label{fig2}
\end{figure*}

Changes to the magnetic state after optical pumping were recorded by imaging the initial (before) and final (after excitation) states, following excitation with 1~MHz laser pulses of photon energy 1.2\,eV, using variable fluence and pulse numbers.
Figure \ref{fig1}b shows the experimental setup, where the initial remanent labyrinth domain state is switched to a skyrmion/bubble domain final state through the application of 125 laser pulses (125 $\mu$s at 1 MHz). This remanent state is not truly a zero-field condition due to a residual field from the electromagnet, estimated to be approximately 5\,mT. In Fig. \ref{fig1}c, a diagram of pump fluence versus the number of pulses required for switching or domain reorientation is shown. This demonstrates that as the pulse fluence is decreased, an increased number of pulses is required to enable switching. The lowest number of pulses required to achieve the switching was 20, at a fluence of 12.6 mJ/cm$^2$. The observation of domain reorientation using multiple pulses points to a heat accumulation effect. We also performed pumping at increased fluence, above 15 mJ/cm$^2$, to achieve single-shot switching. This leads to degradation of the sample, which is generally expected for GaTe-based vdW compounds under ambient conditions \cite{Mercado2019}, especially when considering such high laser fluence. To overcome this, further material optimisation such as encapsulation and pumping with different wavelengths can be undertaken to enable single-shot switching. In addition, the laser-induced heating is expected to be more efficient in thin flakes due to the domain memory effect \cite{dabrowski2025ultrafast}, which is associated with the low thermal conductivity of vdW layers. As a result, magnetic moments lying well below the penetration depth of the optical pump remain intact and help restore the magnetic order in the upper, demagnetised layers. However, as domain size decreases with thickness, we were unable to confirm the effect of thickness on the formation of topological spin textures due to the limited resolution of our optical methods. To verify this thickness dependence, higher-resolution techniques, such as Lorentz transmission electron microscopy \cite{liu2024magnetic}, are required. 

We now demonstrate the protocol used for toggle switching between two distinct states, depicted in Fig. \ref{fig2}. Initially, the labyrinth state is prepared by sweeping the magnetic field from saturation to remanence. The laser is then applied with a small +6 mT field, resulting in a transition to the skyrmion/bubble state. Subsequently, an out-of-plane magnetic field of +\,150 mT is applied, leading to a more distinct formation of skyrmions/bubbles, with the average domain area decreasing from 0.3 $\mu$m$^2$ to 0.2 $\mu$m$^2$ (Fig. \ref{fig2}c), which are close to the resolution limit of the WFKM. Based on previous reports \cite{lv2024distinct,liu2024magnetic,li2024room,Hou2024}, smaller bubbles/skyrmions are likely present in the system but cannot be resolved with our method. By applying laser pulses again to the skyrmion/bubble state at +\,150 mT, we induce a transition back to the labyrinth state (Fig. \ref{fig2}d). Finally, upon removal of the magnetic field, the remanence labyrinth state is recovered, characterised by equally sized and populated up and down domains (Fig. \ref{fig2}e). Notably, in contrast to non-stoichiometric Fe$_{2.84\pm0.05}$GaTe$_2$ \cite{li2024room}, where one-way switching to skyrmions required an external field, we demonstrate two distinct switching events: from the labyrinth state to skyrmions and back to the labyrinth state, via optical pumping at remanence and under an external field, respectively.

To gain deeper insight into the mechanism behind this switching behaviour, we performed micromagnetic simulations using MuMax3 \cite{vansteenkiste2014design}. Using representative values from the literature for the exchange stiffness $A_\mathrm{ex}$, magnetic anisotropy energy, saturation magnetisation, and DMI, we simulate a geometry of $1\,\mu\mathrm{m} \times 1\,\mu\mathrm{m} \times 50\,\mathrm{nm}$  \cite{liu2024magnetic} (see Methods for details). To reproduce the laser-induced heating, the system was evolved for 20~ns at fixed temperature steps between 300~K and 370~K.

\begin{figure*}[!h]
    \centering
    \includegraphics[width=0.8\linewidth]{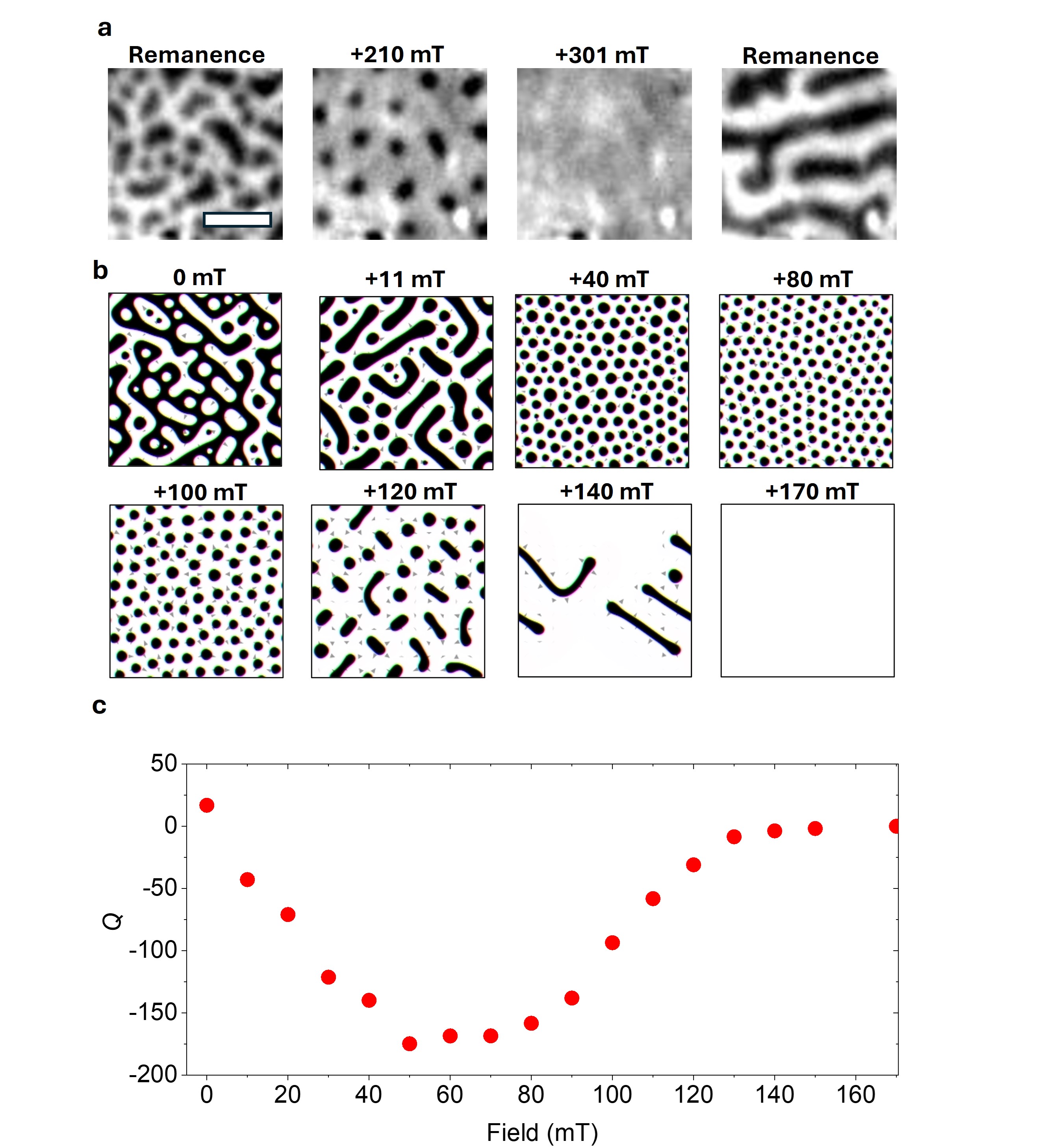}
    \caption{(a) Effect of the applied magnetic field on the laser-induced switched state obtained at remanence after optical pumping for 1500 $\mu$s at 70 mW. The scale bar is 2~$\mu$m. (b) Micromagnetic simulations replicating laser-induced heating under different magnetic field values. The simulation size is 1~$\mu$m~$\times$~1~$\mu$m~$\times$~50~nm. (c) Final topological charge as a function of applied field calculated from the simulations in (b).}

    \label{fig:enter-label}
\end{figure*}


We closely reproduce the experimental data by applying conditions similar to the experimental process, though the exact agreement with the applied fields is not achieved, and attributed to limitations in the system size in the simulation (Fig. \ref{fig2}f-j) as the true bulk thickness is too large to simulate. The field values chosen here provide the closest match to the experimental data. 

We now demonstrate the field stability of the nucleated skyrmion/bubble domains. In Fig.~3a, it can be seen that upon increasing the field, the skyrmion/bubble domains remain stable up to magnetic fields just below the saturation field. When the field is increased beyond saturation and subsequently reduced back to remanence, the labyrinth domains are recovered.

In Fig.~3b, we present simulation results for various applied field values during which the heating process was applied, in order to replicate laser-induced changes to the magnetic structures. At zero field, a hybrid state of up and down bubbles/stripes is observed, similar to previous results reported in CrGeTe$_3$~\cite{Khela2023}. The application of a small positive field breaks the symmetry of the system, promoting the nucleation of skyrmions/bubbles with a uniform polarity. However, at low fields (+11 mT) there is significant size variation, with many elongated bubbles being stabilised. As the field increases, the number of bubbles domains increases, although notable size variation persists relative to an ideal skyrmion lattice. This is attributed to the role of dipole–dipole interactions competing with DMI at low fields when stabilsing skyrmion/bubble domains~\cite{zhang2024above}.
When a field of 120~mT is applied during laser-induced heating, the skyrmion/bubble domains are more sparsely nucleated, and many domains exhibit elongation. At an applied field of 140~mT, only labyrinthine domains are observed after the laser pulse, indicating that a large field beyond the skyrmion/bubble formation window is required to generate labyrinthine/stripe domains. Finally, at fields above a saturation only a uniform domain state is observed.
The final topological charge is then calculated and shown in Fig.~3c. Here the topological charge, $Q$, is defined as,

\begin{equation}
    Q =\frac{1}{4\pi }\int{{{{{{{\bf{n}}}}}}}}\cdot \left(\frac{\partial {{{{{{{\bf{n}}}}}}}}}{\partial x}\times \frac{\partial {{{{{{{\bf{n}}}}}}}}}{\partial y}\right){d}^{2}{{{{{{{\bf{r}}}}}}}}
\end{equation}

where $\mathbf{n}$ is the direction vector of magnetisation, and a skyrmion has a topological charge of -1 \cite{kim2020quantifying}. The highest number of skyrmions (minimum $Q$) are achieved at an applied field of 50~mT. Beyond this field the topological charge gradually increases to zero, reflecting the reduced skyrmion density and the re-emergence of labyrinth domains before the uniform domain above the saturation field. This behaviour demonstrates that the number of skyrmions, can be finely tuned via the applied magnetic field during the switching process. 
Previous studies employing slow field-cooling procedures on \ce{Fe3GaTe2}~\cite{yin2025manipulation,liu2024magnetic,mi2024real} have shown the stabilisation of skyrmion/bubble domains within a field window with a lower boundary of 32~mT and an upper boundary of 55~mT. Considering this and the results shown in Fig.~3b, the observed topological switching behavior can be explained as a laser-induced heating and cooling where the applied field determines the domain state stabilised. When the laser fluence is sufficient to heat the sample above the $T_\mathrm{C}$, the metastable state is erased, and a new domain configuration is established during fast cooling following the removal of the laser. This fast cooling appears to lower the field required for skyrmion/bubble domain formation compared to conventional slow field-cooling procedures, though further comparative experiments should be undertaken to probe the role of fast cooling. In this study, we demonstrate skyrmion/bubble nucleation at approximately 6~mT, representing a five-fold reduction in the applied field required.

\section*{Conclusions}
In summary, we have demonstrated reversible topological toggle switching  between two distinct domain states, the labyrinth and skyrmion/bubble state, through laser-induced heating at room temperature in the bulk vdW ferromagnet \ce{Fe3GaTe2}. The physical process behind this is confirmed through fast heating and quenching in micromagnetic simulations. Beyond these two states, there are opportunities to explore optical control across a rich landscape of possible topological spin configurations found in vdW ferromagnets. 

\section{Methods}
\subsection*{Material Growth}
\ce{Fe3GaTe2} crystals were synthesised by a self-flux method to preclude any possibility of extrinsic doping and contamination. High-purity Fe powder (99.9 \%, Thermo-fisher scientific) and Te powder (99.999 \%, 5N-plus) were ground in a stoichiometric ratio and placed in the quartz tube with Ga granules (99.99999 \%, Alfa-aesar). The quartz tube was sealed under a high vacuum ($<$ 1×10-5 hPa) and heated up to 1373 K within 5 hours. The temperature was held for for 30 hours to ensure the melting of mixtures, and the tube was rapidly cooled down to 1103 K. After the cooldown, the temperature was maintained over 150 hours, and the quartz tube was quenched in air. Single crystals were then separatedfrom the residual flux by mechanical cleaving and exfoliation with Scotch tape so that plate-shaped crystals with typical size of 5 mm×5 mm×0.1 mm were finally obtained.

\subsection*{Micromagnetic simulations} 
Micromagnetic simulations were performed using MuMax3~\cite{vansteenkiste2014design}. A simulation grid of $200 \times 200 \times 10$ cells with a cell size of $5 \times 5 \times 5$~nm$^3$ was used, along with periodic boundary conditions (10,10,0) Magnetic parameters were scaled phenomenologically according to $[M_s(T)/M_s(300K)]  = [K_u(T)/K_u(300K)]^{1/3} = [A_{ex}(T)/A_{ex}(300K)]^{1/2} = [D(T)/D(300K)]^{1/2} $ \cite{richter2018temperature,grebenchuk2024topological}, where $M_s (300K) = 3.76 \times10^{-5} A/m$, $K_u (300K) = 2\times10^{5} J/m^3$, $A_{ex} (300K) = 7.5 \times10^{-12} J/m$, $D (300K) = 1.6 \times10^{-3} J/m^2$. 
We assume that $M_\mathrm{s}(T)$ varies linearly in this temperature region; however, further improvements could be made by using experimental $M_\mathrm{s}(T)$ data. Magnetic parameters were taken from the literature, with the DMI being chosen for good visual match to experimental data \cite{liu2024magnetic,zhang2022above,li2024room}. The damping constant was set to 0.8 to reduce simulation time. The initial labyrinth domain state was prepared by zero-field cooling from a randomised magnetisation configuration, followed by energy minimisation. To emulate laser-induced heating, the system was evolved for 20~ns at selected temperatures between 300~K and 370~K. Temperature effects were incorporated by scaling the magnetic parameters according to the corresponding temperature.

\subsection*{Data Availability} 
All relevant data are available from the corresponding author upon reasonable request.
\subsection{Acknowledgements}
 C.W.F.F. thanks EPSRC for support through EPSRC DTP Case studentship (EP/T517793/1). C.W.F.F. and M.C. acknowledge support from the UK Government Department for Science, Innovation and Technology through National Measurement System. H.K. thanks the Leverhulme Trust for financial support via their Research Fellowship (RF-2024-317) and acknowledges support from EPSRC via EP/T006749/1, EP/V035630/1 and EP/X015661/1. E.J.G.S. acknowledges computational resources through CIRRUS Tier-2 HPC Service (ec131 Cirrus Project) at EPCC (http://www.cirrus.ac.uk) funded by the University of Edinburgh and EPSRC (EP/P020267/1); and ARCHER2 UK National Supercomputing Service via the UKCP consortium (Project e89) funded by EPSRC grant ref EP/X035891/1. E.J.G.S. acknowledges the EPSRC Open Fellowship (EP/T021578/1) and the Donostia International Physics Center for funding support. P.C. acknowledges funding support from the China Scholarship Council grant (202208060246).
\bibliography{main.bib}

\providecommand{\latin}[1]{#1}
\makeatletter
\providecommand{\doi}
  {\begingroup\let\do\@makeother\dospecials
  \catcode`\{=1 \catcode`\}=2 \doi@aux}
\providecommand{\doi@aux}[1]{\endgroup\texttt{#1}}
\makeatother
\providecommand*\mcitethebibliography{\thebibliography}
\csname @ifundefined\endcsname{endmcitethebibliography}  {\let\endmcitethebibliography\endthebibliography}{}
\begin{mcitethebibliography}{39}
\providecommand*\natexlab[1]{#1}
\providecommand*\mciteSetBstSublistMode[1]{}
\providecommand*\mciteSetBstMaxWidthForm[2]{}
\providecommand*\mciteBstWouldAddEndPuncttrue
  {\def\EndOfBibitem{\unskip.}}
\providecommand*\mciteBstWouldAddEndPunctfalse
  {\let\EndOfBibitem\relax}
\providecommand*\mciteSetBstMidEndSepPunct[3]{}
\providecommand*\mciteSetBstSublistLabelBeginEnd[3]{}
\providecommand*\EndOfBibitem{}
\mciteSetBstSublistMode{f}
\mciteSetBstMaxWidthForm{subitem}{(\alph{mcitesubitemcount})}
\mciteSetBstSublistLabelBeginEnd
  {\mcitemaxwidthsubitemform\space}
  {\relax}
  {\relax}

\bibitem[Gong \latin{et~al.}(2017)Gong, Li, Li, Ji, Stern, Xia, Cao, Bao, Wang, Wang, \latin{et~al.} others]{gong2017discovery}
Gong,~C.; Li,~L.; Li,~Z.; Ji,~H.; Stern,~A.; Xia,~Y.; Cao,~T.; Bao,~W.; Wang,~C.; Wang,~Y.; others Discovery of intrinsic ferromagnetism in two-dimensional van der Waals crystals. \emph{Nature} \textbf{2017}, \emph{546}, 265--269\relax
\mciteBstWouldAddEndPuncttrue
\mciteSetBstMidEndSepPunct{\mcitedefaultmidpunct}
{\mcitedefaultendpunct}{\mcitedefaultseppunct}\relax
\EndOfBibitem
\bibitem[Huang \latin{et~al.}(2017)Huang, Clark, Navarro-Moratalla, Klein, Cheng, Seyler, Zhong, Schmidgall, McGuire, Cobden, \latin{et~al.} others]{huang2017layer}
Huang,~B.; Clark,~G.; Navarro-Moratalla,~E.; Klein,~D.~R.; Cheng,~R.; Seyler,~K.~L.; Zhong,~D.; Schmidgall,~E.; McGuire,~M.~A.; Cobden,~D.~H.; others Layer-dependent ferromagnetism in a van der Waals crystal down to the monolayer limit. \emph{Nature} \textbf{2017}, \emph{546}, 270--273\relax
\mciteBstWouldAddEndPuncttrue
\mciteSetBstMidEndSepPunct{\mcitedefaultmidpunct}
{\mcitedefaultendpunct}{\mcitedefaultseppunct}\relax
\EndOfBibitem
\bibitem[Kurebayashi \latin{et~al.}(2022)Kurebayashi, Garcia, Khan, Sinova, and Roche]{kurebayashi2022magnetism}
Kurebayashi,~H.; Garcia,~J.~H.; Khan,~S.; Sinova,~J.; Roche,~S. Magnetism, symmetry and spin transport in van der Waals layered systems. \emph{Nature Reviews Physics} \textbf{2022}, \emph{4}, 150--166\relax
\mciteBstWouldAddEndPuncttrue
\mciteSetBstMidEndSepPunct{\mcitedefaultmidpunct}
{\mcitedefaultendpunct}{\mcitedefaultseppunct}\relax
\EndOfBibitem
\bibitem[Jiang \latin{et~al.}(2021)Jiang, Liu, Xing, Liu, Guo, Liu, and Zhao]{jiang2021recent}
Jiang,~X.; Liu,~Q.; Xing,~J.; Liu,~N.; Guo,~Y.; Liu,~Z.; Zhao,~J. Recent progress on 2D magnets: Fundamental mechanism, structural design and modification. \emph{Applied Physics Reviews} \textbf{2021}, \emph{8}\relax
\mciteBstWouldAddEndPuncttrue
\mciteSetBstMidEndSepPunct{\mcitedefaultmidpunct}
{\mcitedefaultendpunct}{\mcitedefaultseppunct}\relax
\EndOfBibitem
\bibitem[Gong and Zhang(2019)Gong, and Zhang]{gong2019two}
Gong,~C.; Zhang,~X. Two-dimensional magnetic crystals and emergent heterostructure devices. \emph{Science} \textbf{2019}, \emph{363}, eaav4450\relax
\mciteBstWouldAddEndPuncttrue
\mciteSetBstMidEndSepPunct{\mcitedefaultmidpunct}
{\mcitedefaultendpunct}{\mcitedefaultseppunct}\relax
\EndOfBibitem
\bibitem[Wang \latin{et~al.}(2025)Wang, Graham, Rama-Eiroa, Islam, Badarneh, Nunes~Gontijo, Tiwari, Adhikari, Zhang, Watanabe, \latin{et~al.} others]{wang2025configurable}
Wang,~Y.-X.; Graham,~T.~K.; Rama-Eiroa,~R.; Islam,~M.~A.; Badarneh,~M.~H.; Nunes~Gontijo,~R.; Tiwari,~G.~P.; Adhikari,~T.; Zhang,~X.-Y.; Watanabe,~K.; others Configurable antiferromagnetic domains and lateral exchange bias in atomically thin CrPS4. \emph{Nature Materials} \textbf{2025}, 1--10\relax
\mciteBstWouldAddEndPuncttrue
\mciteSetBstMidEndSepPunct{\mcitedefaultmidpunct}
{\mcitedefaultendpunct}{\mcitedefaultseppunct}\relax
\EndOfBibitem
\bibitem[Burch \latin{et~al.}(2018)Burch, Mandrus, and Park]{burch2018magnetism}
Burch,~K.~S.; Mandrus,~D.; Park,~J.-G. Magnetism in two-dimensional van der Waals materials. \emph{Nature} \textbf{2018}, \emph{563}, 47--52\relax
\mciteBstWouldAddEndPuncttrue
\mciteSetBstMidEndSepPunct{\mcitedefaultmidpunct}
{\mcitedefaultendpunct}{\mcitedefaultseppunct}\relax
\EndOfBibitem
\bibitem[Wang \latin{et~al.}(2022)Wang, Bedoya-Pinto, Blei, Dismukes, Hamo, Jenkins, Koperski, Liu, Sun, Telford, \latin{et~al.} others]{wang2022magnetic}
Wang,~Q.~H.; Bedoya-Pinto,~A.; Blei,~M.; Dismukes,~A.~H.; Hamo,~A.; Jenkins,~S.; Koperski,~M.; Liu,~Y.; Sun,~Q.-C.; Telford,~E.~J.; others The magnetic genome of two-dimensional van der Waals materials. \emph{ACS nano} \textbf{2022}, \emph{16}, 6960--7079\relax
\mciteBstWouldAddEndPuncttrue
\mciteSetBstMidEndSepPunct{\mcitedefaultmidpunct}
{\mcitedefaultendpunct}{\mcitedefaultseppunct}\relax
\EndOfBibitem
\bibitem[Zhou \latin{et~al.}(2025)Zhou, Li, Liang, and Zhou]{Zhou2025}
Zhou,~Y.; Li,~S.; Liang,~X.; Zhou,~Y. Topological Spin Textures: Basic Physics and Devices. \emph{Advanced Materials} \textbf{2025}, \emph{37}, 2312935\relax
\mciteBstWouldAddEndPuncttrue
\mciteSetBstMidEndSepPunct{\mcitedefaultmidpunct}
{\mcitedefaultendpunct}{\mcitedefaultseppunct}\relax
\EndOfBibitem
\bibitem[Berruto \latin{et~al.}(2018)Berruto, Madan, Murooka, Vanacore, Pomarico, Rajeswari, Lamb, Huang, Kruchkov, Togawa, LaGrange, McGrouther, R\o{}nnow, and Carbone]{Berruto2018}
Berruto,~G.; Madan,~I.; Murooka,~Y.; Vanacore,~G.~M.; Pomarico,~E.; Rajeswari,~J.; Lamb,~R.; Huang,~P.; Kruchkov,~A.~J.; Togawa,~Y.; LaGrange,~T.; McGrouther,~D.; R\o{}nnow,~H.~M.; Carbone,~F. Laser-Induced Skyrmion Writing and Erasing in an Ultrafast Cryo-Lorentz Transmission Electron Microscope. \emph{Phys. Rev. Lett.} \textbf{2018}, \emph{120}, 117201\relax
\mciteBstWouldAddEndPuncttrue
\mciteSetBstMidEndSepPunct{\mcitedefaultmidpunct}
{\mcitedefaultendpunct}{\mcitedefaultseppunct}\relax
\EndOfBibitem
\bibitem[Gerlinger \latin{et~al.}(2021)Gerlinger, Pfau, Büttner, Schneider, Kern, Fuchs, Engel, Günther, Huang, Lemesh, Caretta, Churikova, Hessing, Klose, Strüber, Schmising, Huang, Wittmann, Litzius, Metternich, Battistelli, Bagschik, Sadovnikov, Beach, and Eisebitt]{Gerlinger2021}
Gerlinger,~K. \latin{et~al.}  Application concepts for ultrafast laser-induced skyrmion creation and annihilation. \emph{Applied Physics Letters} \textbf{2021}, \emph{118}, 192403\relax
\mciteBstWouldAddEndPuncttrue
\mciteSetBstMidEndSepPunct{\mcitedefaultmidpunct}
{\mcitedefaultendpunct}{\mcitedefaultseppunct}\relax
\EndOfBibitem
\bibitem[Haldar \latin{et~al.}(2025)Haldar, Griepe, Atxitia, and Santos]{haldar2025all}
Haldar,~S.; Griepe,~T.; Atxitia,~U.; Santos,~E.~J. All-Heat Control of Magnetization Dynamics on Van der Waals Magnets. \emph{Advanced Materials} \textbf{2025}, 2501043\relax
\mciteBstWouldAddEndPuncttrue
\mciteSetBstMidEndSepPunct{\mcitedefaultmidpunct}
{\mcitedefaultendpunct}{\mcitedefaultseppunct}\relax
\EndOfBibitem
\bibitem[Kern \latin{et~al.}(2025)Kern, Kuchkin, Deinhart, Klose, Sidiropoulos, Auer, Gaebel, Gerlinger, Battistelli, Wittrock, Karaman, Schneider, Günther, Engel, Will, Wintz, Weigand, Büttner, Höflich, Eisebitt, and Pfau]{Kern2025}
Kern,~L.-M. \latin{et~al.}  Controlled Formation of Skyrmion Bags. \emph{Advanced Materials} \textbf{2025}, \emph{n/a}, 2501250\relax
\mciteBstWouldAddEndPuncttrue
\mciteSetBstMidEndSepPunct{\mcitedefaultmidpunct}
{\mcitedefaultendpunct}{\mcitedefaultseppunct}\relax
\EndOfBibitem
\bibitem[Beaurepaire \latin{et~al.}(1996)Beaurepaire, Merle, Daunois, and Bigot]{Beaurepaire1996}
Beaurepaire,~E.; Merle,~J.-C.; Daunois,~A.; Bigot,~J.-Y. Ultrafast Spin Dynamics in Ferromagnetic Nickel. \emph{Phys. Rev. Lett.} \textbf{1996}, \emph{76}, 4250--4253\relax
\mciteBstWouldAddEndPuncttrue
\mciteSetBstMidEndSepPunct{\mcitedefaultmidpunct}
{\mcitedefaultendpunct}{\mcitedefaultseppunct}\relax
\EndOfBibitem
\bibitem[Khela \latin{et~al.}(2023)Khela, D\k{a}browski, Khan, Keatley, Verzhbitskiy, Eda, Hicken, Kurebayashi, and Santos]{Khela2023}
Khela,~M.; D\k{a}browski,~M.; Khan,~S.; Keatley,~P.~S.; Verzhbitskiy,~I.; Eda,~G.; Hicken,~R.~J.; Kurebayashi,~H.; Santos,~E. J.~G. Laser-induced topological spin switching in a 2D van der Waals magnet. \emph{Nature Communications} \textbf{2023}, \emph{14}, 1378\relax
\mciteBstWouldAddEndPuncttrue
\mciteSetBstMidEndSepPunct{\mcitedefaultmidpunct}
{\mcitedefaultendpunct}{\mcitedefaultseppunct}\relax
\EndOfBibitem
\bibitem[Lambert \latin{et~al.}(2014)Lambert, Mangin, Varaprasad, Takahashi, Hehn, Cinchetti, Malinowski, Hono, Fainman, Aeschlimann, and Fullerton]{Lambert2014}
Lambert,~C.-H.; Mangin,~S.; Varaprasad,~B. S. D. C.~S.; Takahashi,~Y.~K.; Hehn,~M.; Cinchetti,~M.; Malinowski,~G.; Hono,~K.; Fainman,~Y.; Aeschlimann,~M.; Fullerton,~E.~E. All-optical control of ferromagnetic thin films and nanostructures. \emph{Science} \textbf{2014}, \emph{345}, 1337--1340\relax
\mciteBstWouldAddEndPuncttrue
\mciteSetBstMidEndSepPunct{\mcitedefaultmidpunct}
{\mcitedefaultendpunct}{\mcitedefaultseppunct}\relax
\EndOfBibitem
\bibitem[Stanciu \latin{et~al.}(2007)Stanciu, Hansteen, Kimel, Kirilyuk, Tsukamoto, Itoh, and Rasing]{Stanciu2007}
Stanciu,~C.~D.; Hansteen,~F.; Kimel,~A.~V.; Kirilyuk,~A.; Tsukamoto,~A.; Itoh,~A.; Rasing,~T. All-Optical Magnetic Recording with Circularly Polarized Light. \emph{Phys. Rev. Lett.} \textbf{2007}, \emph{99}, 047601\relax
\mciteBstWouldAddEndPuncttrue
\mciteSetBstMidEndSepPunct{\mcitedefaultmidpunct}
{\mcitedefaultendpunct}{\mcitedefaultseppunct}\relax
\EndOfBibitem
\bibitem[D\k{a}browski \latin{et~al.}(2021)D\k{a}browski, Scott, Hendren, Forbes, Frisk, Burn, Newman, Sait, Keatley, N’Diaye, Hesjedal, van~der Laan, Bowman, and Hicken]{Dabrowski2021}
D\k{a}browski,~M.; Scott,~J.~N.; Hendren,~W.~R.; Forbes,~C.~M.; Frisk,~A.; Burn,~D.~M.; Newman,~D.~G.; Sait,~C. R.~J.; Keatley,~P.~S.; N’Diaye,~A.~T.; Hesjedal,~T.; van~der Laan,~G.; Bowman,~R.~M.; Hicken,~R.~J. Transition Metal Synthetic Ferrimagnets: Tunable Media for All-Optical Switching Driven by Nanoscale Spin Current. \emph{Nano Lett.} \textbf{2021}, \emph{21}, 9210--9216\relax
\mciteBstWouldAddEndPuncttrue
\mciteSetBstMidEndSepPunct{\mcitedefaultmidpunct}
{\mcitedefaultendpunct}{\mcitedefaultseppunct}\relax
\EndOfBibitem
\bibitem[Zhang \latin{et~al.}(2022)Zhang, Chung, Li, Wang, Wang, Huey, Yang, Goldberger, Yao, and Zhang]{Zhang2022a}
Zhang,~P.; Chung,~T.-F.; Li,~Q.; Wang,~S.; Wang,~Q.; Huey,~W. L.~B.; Yang,~S.; Goldberger,~J.~E.; Yao,~J.; Zhang,~X. All-optical switching of magnetization in atomically thin CrI3. \emph{Nature Materials} \textbf{2022}, \relax
\mciteBstWouldAddEndPunctfalse
\mciteSetBstMidEndSepPunct{\mcitedefaultmidpunct}
{}{\mcitedefaultseppunct}\relax
\EndOfBibitem
\bibitem[Dabrowski \latin{et~al.}(2022)Dabrowski, Guo, Strungaru, Keatley, Withers, Santos, and Hicken]{Dabrowski2022}
Dabrowski,~M.; Guo,~S.; Strungaru,~M.; Keatley,~P.~S.; Withers,~F.; Santos,~E.~J.; Hicken,~R.~J. All-optical control of spin in a 2D van der Waals magnet. \emph{Nature Communications} \textbf{2022}, \emph{13}, 5976\relax
\mciteBstWouldAddEndPuncttrue
\mciteSetBstMidEndSepPunct{\mcitedefaultmidpunct}
{\mcitedefaultendpunct}{\mcitedefaultseppunct}\relax
\EndOfBibitem
\bibitem[Zhang \latin{et~al.}(2022)Zhang, Guo, Wu, Wen, Yang, Jin, Zhang, and Chang]{zhang2022above}
Zhang,~G.; Guo,~F.; Wu,~H.; Wen,~X.; Yang,~L.; Jin,~W.; Zhang,~W.; Chang,~H. Above-room-temperature strong intrinsic ferromagnetism in 2D van der Waals \ce{Fe3GaTe2} with large perpendicular magnetic anisotropy. \emph{Nature Communications} \textbf{2022}, \emph{13}, 5067\relax
\mciteBstWouldAddEndPuncttrue
\mciteSetBstMidEndSepPunct{\mcitedefaultmidpunct}
{\mcitedefaultendpunct}{\mcitedefaultseppunct}\relax
\EndOfBibitem
\bibitem[Lv \latin{et~al.}(2024)Lv, Lv, Huang, Zhang, Qin, Dong, Liu, Pei, Cao, Zhang, \latin{et~al.} others]{lv2024distinct}
Lv,~X.; Lv,~H.; Huang,~Y.; Zhang,~R.; Qin,~G.; Dong,~Y.; Liu,~M.; Pei,~K.; Cao,~G.; Zhang,~J.; others Distinct skyrmion phases at room temperature in two-dimensional ferromagnet Fe3GaTe2. \emph{Nature Communications} \textbf{2024}, \emph{15}, 3278\relax
\mciteBstWouldAddEndPuncttrue
\mciteSetBstMidEndSepPunct{\mcitedefaultmidpunct}
{\mcitedefaultendpunct}{\mcitedefaultseppunct}\relax
\EndOfBibitem
\bibitem[Hu \latin{et~al.}(2024)Hu, Guo, Lv, Li, Wang, Han, Pan, Xie, Yu, Zhu, \latin{et~al.} others]{hu2024room}
Hu,~G.; Guo,~H.; Lv,~S.; Li,~L.; Wang,~Y.; Han,~Y.; Pan,~L.; Xie,~Y.; Yu,~W.; Zhu,~K.; others Room-Temperature Antisymmetric Magnetoresistance in van der Waals Ferromagnet Fe3GaTe2 Nanosheets. \emph{Advanced Materials} \textbf{2024}, \emph{36}, 2403154\relax
\mciteBstWouldAddEndPuncttrue
\mciteSetBstMidEndSepPunct{\mcitedefaultmidpunct}
{\mcitedefaultendpunct}{\mcitedefaultseppunct}\relax
\EndOfBibitem
\bibitem[Ruiz \latin{et~al.}(2024)Ruiz, Esteras, L{\'o}pez-Alcal{\'a}, and Baldov{\'\i}]{ruiz2024origin}
Ruiz,~A.~M.; Esteras,~D.~L.; L{\'o}pez-Alcal{\'a},~D.; Baldov{\'\i},~J.~J. On the origin of the above-room-temperature magnetism in the 2D van der Waals Ferromagnet Fe3GaTe2. \emph{Nano Letters} \textbf{2024}, \emph{24}, 7886--7894\relax
\mciteBstWouldAddEndPuncttrue
\mciteSetBstMidEndSepPunct{\mcitedefaultmidpunct}
{\mcitedefaultendpunct}{\mcitedefaultseppunct}\relax
\EndOfBibitem
\bibitem[Liu \latin{et~al.}(2024)Liu, Zhang, Hao, Algaidi, Ma, and Zhang]{liu2024magnetic}
Liu,~C.; Zhang,~S.; Hao,~H.; Algaidi,~H.; Ma,~Y.; Zhang,~X.-X. Magnetic skyrmions above room temperature in a van der Waals ferromagnet Fe3GaTe2. \emph{Advanced Materials} \textbf{2024}, \emph{36}, 2311022\relax
\mciteBstWouldAddEndPuncttrue
\mciteSetBstMidEndSepPunct{\mcitedefaultmidpunct}
{\mcitedefaultendpunct}{\mcitedefaultseppunct}\relax
\EndOfBibitem
\bibitem[Li \latin{et~al.}(2024)Li, Zhang, Li, Guo, Wang, Deng, Hu, Hu, Liu, Qin, \latin{et~al.} others]{li2024room}
Li,~Z.; Zhang,~H.; Li,~G.; Guo,~J.; Wang,~Q.; Deng,~Y.; Hu,~Y.; Hu,~X.; Liu,~C.; Qin,~M.; others Room-temperature sub-100 nm N{\'e}el-type skyrmions in non-stoichiometric van der Waals ferromagnet Fe3-x GaTe2 with ultrafast laser writability. \emph{Nature Communications} \textbf{2024}, \emph{15}, 1017\relax
\mciteBstWouldAddEndPuncttrue
\mciteSetBstMidEndSepPunct{\mcitedefaultmidpunct}
{\mcitedefaultendpunct}{\mcitedefaultseppunct}\relax
\EndOfBibitem
\bibitem[Zhang \latin{et~al.}(2024)Zhang, Jiang, Jiang, He, Zhang, Hu, Zhao, Yang, Liu, Peng, \latin{et~al.} others]{zhang2024above}
Zhang,~C.; Jiang,~Z.; Jiang,~J.; He,~W.; Zhang,~J.; Hu,~F.; Zhao,~S.; Yang,~D.; Liu,~Y.; Peng,~Y.; others Above-room-temperature chiral skyrmion lattice and Dzyaloshinskii--Moriya interaction in a van der Waals ferromagnet Fe3- x GaTe2. \emph{Nature Communications} \textbf{2024}, \emph{15}, 4472\relax
\mciteBstWouldAddEndPuncttrue
\mciteSetBstMidEndSepPunct{\mcitedefaultmidpunct}
{\mcitedefaultendpunct}{\mcitedefaultseppunct}\relax
\EndOfBibitem
\bibitem[Luo \latin{et~al.}(2025)Luo, Fang, Miao, Zhao, Shen, Fan, Ma, Wang, Shen, Zhang, \latin{et~al.} others]{luo2025manipulation}
Luo,~X.; Fang,~M.; Miao,~L.-P.; Zhao,~J.; Shen,~Z.; Fan,~W.; Ma,~X.; Wang,~Z.; Shen,~A.; Zhang,~J.; others Manipulation of magnetic spin textures at room temperature in a van der Waals ferromagnet. \emph{Physical Review B} \textbf{2025}, \emph{111}, 014442\relax
\mciteBstWouldAddEndPuncttrue
\mciteSetBstMidEndSepPunct{\mcitedefaultmidpunct}
{\mcitedefaultendpunct}{\mcitedefaultseppunct}\relax
\EndOfBibitem
\bibitem[Mi \latin{et~al.}(2024)Mi, Guo, Hu, Wang, Li, Gong, Jin, Xu, Pang, Ji, \latin{et~al.} others]{mi2024real}
Mi,~S.; Guo,~J.; Hu,~G.; Wang,~G.; Li,~S.; Gong,~Z.; Jin,~S.; Xu,~R.; Pang,~F.; Ji,~W.; others Real-space topology-engineering of skyrmionic spin textures in a van der Waals ferromagnet Fe3GaTe2. \emph{Nano Letters} \textbf{2024}, \emph{24}, 13094--13102\relax
\mciteBstWouldAddEndPuncttrue
\mciteSetBstMidEndSepPunct{\mcitedefaultmidpunct}
{\mcitedefaultendpunct}{\mcitedefaultseppunct}\relax
\EndOfBibitem
\bibitem[Steger(1998)]{steger1998unbiased}
Steger,~C. An unbiased detector of curvilinear structures. \emph{IEEE Transactions on pattern analysis and machine intelligence} \textbf{1998}, \emph{20}, 113--125\relax
\mciteBstWouldAddEndPuncttrue
\mciteSetBstMidEndSepPunct{\mcitedefaultmidpunct}
{\mcitedefaultendpunct}{\mcitedefaultseppunct}\relax
\EndOfBibitem
\bibitem[Mercado \latin{et~al.}(2019)Mercado, Zhou, Xie, Zhao, Cai, Chen, Jie, Tongay, Wang, and Kuball]{Mercado2019}
Mercado,~E.; Zhou,~Y.; Xie,~Y.; Zhao,~Q.; Cai,~H.; Chen,~B.; Jie,~W.; Tongay,~S.; Wang,~T.; Kuball,~M. Passivation of Layered Gallium Telluride by Double Encapsulation with Graphene. \emph{ACS Omega} \textbf{2019}, \emph{4}, 18002--18010\relax
\mciteBstWouldAddEndPuncttrue
\mciteSetBstMidEndSepPunct{\mcitedefaultmidpunct}
{\mcitedefaultendpunct}{\mcitedefaultseppunct}\relax
\EndOfBibitem
\bibitem[Dabrowski \latin{et~al.}(2025)Dabrowski, Haldar, Khan, Keatley, Sagkovits, Xue, Freeman, Verzhbitskiy, Griepe, Atxitia, \latin{et~al.} others]{dabrowski2025ultrafast}
Dabrowski,~M.; Haldar,~S.; Khan,~S.; Keatley,~P.~S.; Sagkovits,~D.; Xue,~Z.; Freeman,~C.; Verzhbitskiy,~I.; Griepe,~T.; Atxitia,~U.; others Ultrafast thermo-optical control of spins in a 2D van der Waals semiconductor. \emph{Nature Communications} \textbf{2025}, \emph{16}, 2797\relax
\mciteBstWouldAddEndPuncttrue
\mciteSetBstMidEndSepPunct{\mcitedefaultmidpunct}
{\mcitedefaultendpunct}{\mcitedefaultseppunct}\relax
\EndOfBibitem
\bibitem[Hou \latin{et~al.}(2024)Hou, Wang, Zhang, Xu, Sun, Li, Wang, Qu, Wei, and Guo]{Hou2024}
Hou,~X.; Wang,~H.; Zhang,~B.; Xu,~C.; Sun,~L.; Li,~Z.; Wang,~X.; Qu,~K.; Wei,~Y.; Guo,~Y. {Room-temperature skyrmions in the van der Waals ferromagnet Fe3GaTe2}. \emph{Applied Physics Letters} \textbf{2024}, \emph{124}, 142404\relax
\mciteBstWouldAddEndPuncttrue
\mciteSetBstMidEndSepPunct{\mcitedefaultmidpunct}
{\mcitedefaultendpunct}{\mcitedefaultseppunct}\relax
\EndOfBibitem
\bibitem[Vansteenkiste \latin{et~al.}(2014)Vansteenkiste, Leliaert, Dvornik, Helsen, Garcia-Sanchez, and Van~Waeyenberge]{vansteenkiste2014design}
Vansteenkiste,~A.; Leliaert,~J.; Dvornik,~M.; Helsen,~M.; Garcia-Sanchez,~F.; Van~Waeyenberge,~B. The design and verification of MuMax3. \emph{AIP advances} \textbf{2014}, \emph{4}\relax
\mciteBstWouldAddEndPuncttrue
\mciteSetBstMidEndSepPunct{\mcitedefaultmidpunct}
{\mcitedefaultendpunct}{\mcitedefaultseppunct}\relax
\EndOfBibitem
\bibitem[Kim and Mulkers(2020)Kim, and Mulkers]{kim2020quantifying}
Kim,~J.-V.; Mulkers,~J. On quantifying the topological charge in micromagnetics using a lattice-based approach. \emph{IOP SciNotes} \textbf{2020}, \emph{1}, 025211\relax
\mciteBstWouldAddEndPuncttrue
\mciteSetBstMidEndSepPunct{\mcitedefaultmidpunct}
{\mcitedefaultendpunct}{\mcitedefaultseppunct}\relax
\EndOfBibitem
\bibitem[Yin \latin{et~al.}(2025)Yin, Liu, Li, Li, Zou, Wu, Xia, and Wang]{yin2025manipulation}
Yin,~Y.; Liu,~M.; Li,~W.; Li,~W.; Zou,~M.; Wu,~S.; Xia,~W.; Wang,~B. Manipulation of room-temperature magnetic skyrmions in a van der Waals ferromagnet Fe3GaTe2. \emph{New Journal of Physics} \textbf{2025}, \relax
\mciteBstWouldAddEndPunctfalse
\mciteSetBstMidEndSepPunct{\mcitedefaultmidpunct}
{}{\mcitedefaultseppunct}\relax
\EndOfBibitem
\bibitem[Richter \latin{et~al.}(2018)Richter, Weber, Martin, Singh, Schwingenschl{\"o}gl, Lotsch, and Kl{\"a}ui]{richter2018temperature}
Richter,~N.; Weber,~D.; Martin,~F.; Singh,~N.; Schwingenschl{\"o}gl,~U.; Lotsch,~B.~V.; Kl{\"a}ui,~M. Temperature-dependent magnetic anisotropy in the layered magnetic semiconductors Cr I 3 and CrB r 3. \emph{Physical Review Materials} \textbf{2018}, \emph{2}, 024004\relax
\mciteBstWouldAddEndPuncttrue
\mciteSetBstMidEndSepPunct{\mcitedefaultmidpunct}
{\mcitedefaultendpunct}{\mcitedefaultseppunct}\relax
\EndOfBibitem
\bibitem[Grebenchuk \latin{et~al.}(2024)Grebenchuk, McKeever, Grzeszczyk, Chen, {\v{S}}i{\v{s}}kins, McCray, Li, Petford-Long, Phatak, Ruihuan, \latin{et~al.} others]{grebenchuk2024topological}
Grebenchuk,~S.; McKeever,~C.; Grzeszczyk,~M.; Chen,~Z.; {\v{S}}i{\v{s}}kins,~M.; McCray,~A.~R.; Li,~Y.; Petford-Long,~A.~K.; Phatak,~C.~M.; Ruihuan,~D.; others Topological spin textures in an insulating van der Waals ferromagnet. \emph{Advanced Materials} \textbf{2024}, \emph{36}, 2311949\relax
\mciteBstWouldAddEndPuncttrue
\mciteSetBstMidEndSepPunct{\mcitedefaultmidpunct}
{\mcitedefaultendpunct}{\mcitedefaultseppunct}\relax
\EndOfBibitem
\end{mcitethebibliography}

\end{document}